\documentclass[journal]{IEEEtran}
\usepackage{color}
\usepackage{threeparttable}
\usepackage{srcltx}
\usepackage{psfrag}
\usepackage{subfigure}
\usepackage{epsfig}
\usepackage{amsmath}

\begin{document}

\title{Single-Order Transmission Diffraction Gratings based on Dispersion Engineered \\All-dielectric Metasurfaces}
\author{Shulabh~Gupta
\thanks{S. Gupta is with the Department
of Electronics, Carleton University, Ottawa, Ontario, Canada e-mail: shulabh.gupta@carleton.ca}
}

\markboth{}%
{Shell \MakeLowercase{\textit{et al.}}: Bare Demo of IEEEtran.cls for IEEE Journals}

\maketitle

\begin{abstract}
A single-order transmission diffraction grating based on dispersion engineered all-dielectric metasurfaces is proposed and its wavelength discriminating properties have been theoretically described and confirmed using numerical simulations. The metasurface is designed using a 2D array of all-dielectric resonators, which emulates a Huygens' source configuration to achieve a perfect match to free-space in broad bandwidth. Using a holey dielectric nanodisk structure as the unit cell, the resonant wavelength is tapered across the metasurface to engineer the wavelength dependent spatial phase gradient, to emulate a dispersive prism. Consequently, different wavelengths are steered towards different directions and thus are discriminated on the output image plane. Due to sub-wavelength periodicities involved, the wavelength discrimination is achieved directly in the zeroth diffraction order of the device, unlike conventional diffraction gratings, thereby providing a high efficiency wavelength discriminating device. 
\end{abstract}

\section{Introduction}

Wavelength discrimination is ubiquitous across the entire electromagnetic spectrum and has widespread applications in communication, instrumentation and ultrafast signal processing applications, such as in spectroscopy \cite{Spectroscopy}, imaging \cite{Spatial_Disperser}, wavelength multiplexing and demultiplexing \cite{DemuxWeiner}\cite{SA_Application1_Jalali}, to name a few.

While an optical prism is conceptually the most basic wavelength discriminator, the commonly used device in free-space optics is a diffraction grating. A diffraction grating is an optical component with a periodic structure, which splits and diffracts light into several beams travelling in different directions \cite{Goodman_Fourier_Optics}. Among the several space harmonics that are generated, except the zeroth order, all diffractions order exhibit wavelength separation, among which only one of the orders is typically used. The rest of the space harmonics including, in particular, the zeroth order, are the unwanted beams and are the prime reasons for lower efficiencies. While some specialized gratings, such as the blazed gratings, maximize the power in a given order, the achievable efficiency in a chosen order can never be unity due to the unavoidable presence of the zeroth order \cite{BlazedGrating}. An excellent alternative to diffraction gratings, is a virtual image phased array (VIPA) device \cite{VIPA_Fujitsu}\cite{VIPA_Disperser}. While they provide fine wavelength resolutions, they exhibit small free-spectral ranges resulting in small bandwidth of operation. More importantly, they always require a collimating cylindrical lens to couple the incident beam inside the structure. Arrayed-waveguide grating (AWGs) is another wavelength separating device, however they are restricted to integrated optical devices only \cite{Saleh_Teich_FP}.

In this work, a wavelength separating free-space device based on artificially engineered electromagnetic structures, or metamaterials, is proposed. It discriminates wavelengths in the transmission mode, without generating spurious diffraction orders, or requiring complex feeding setup. The proposed device is based on metasurfaces, which are two dimensional equivalent of volumetric metamaterials \cite{meta1}\cite{meta2} and have been recently explored for the novel paradigm of flat optical systems \cite{meta3}\cite{RingsMetasurfaceReview}. While they have been mostly focussed on single wavelength operations, they have also been recently applied towards multi-wavelength broadband devices realizing achromatic lenses \cite{Achromatic_MS_Capasso} and wavelength separators in reflection mode \cite{SA_Reflection}\cite{NanoRingSteering}\cite{Zhu_SingleOrder}, to name a few. As opposed to the previous works, this work proposes the wavelength separation in the \emph{transmission mode}. Since light can pass through the grating, transmission gratings can be used in compact in-line configurations and thus are ideal for usage in spectrometers and beam splitters in demultiplexing applications. The transmission mode operation thus requires a constraint on the impedance matching of the grating in a wide bandwidth, in addition to maximizing the grating efficiency.

The paper is structured as follows. Sec. 2 summarizes the basic properties of a transmission diffraction grating taking a thin sinusoidal grating as an example. It further explains the principle of the proposed metasurface gratings and provide the theoretical background to construct metasurface grating using Huygens' unit cells. Sec. 3, provides detailed design of the all-dielectric metasurfaces and shows numerical simulation results to confirm the operation of the grating. Sec 4 summarizes some features and benefits of the proposed metasurface grating and finally, conclusions are provided in Sec. 5.

\section{Principle}

\subsection{Conventional Diffraction Grating}

Let us consider a conventional thin sinusoidal grating with its spatial profile given by

\begin{equation}
g_t(x)=\frac{1}{2}\left[1+m\cos\left(2\pi\frac{x}{\Lambda}\right)\right]
\end{equation}

\noindent where $m$ is the modulation depth and $\Lambda$ is the period of the diffraction grating. When such a grating is illuminated with a plane-wave followed by a focussing lens of focal length $d$, the field amplitude in the focal plane of the lens is given by \cite{Goodman_Fourier_Optics}

\begin{align}
|I(x,y; d)| \propto \left[\frac{\delta(x)}{2} +  \frac{m}{2} \left\{\delta\left(x+  \frac{\lambda d}{\Lambda}\right) + \delta\left(x- \frac{\lambda d}{\Lambda}\right)\right\}\right].
\end{align}

\begin{figure}[htbp]
\begin{center}
\subfigure[]{
\psfrag{a}[c][c][0.8]{\shortstack{source\\ plane}}
\psfrag{b}[c][c][0.8]{\shortstack{Conventional\\ grating}}
\psfrag{c}[c][c][0.8]{$+1^\text{st}$}
\psfrag{d}[c][c][0.8]{$-1^\text{st}$}
\psfrag{e}[c][c][0.8]{$0^\text{th}$}
\psfrag{f}[c][c][0.8]{$\ell_0$}
\psfrag{g}[c][c][0.8]{\shortstack{focal\\ length}}
\psfrag{x}[c][c][0.8]{$x$}
\psfrag{y}[c][c][0.8]{$y$}
\psfrag{z}[c][c][0.8]{$z$}
\psfrag{n}[c][c][0.8]{$\Lambda$}
\includegraphics[width=0.7\columnwidth]{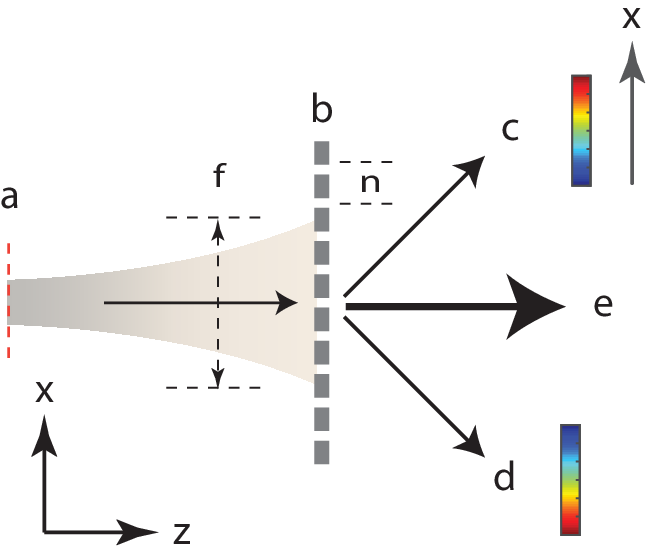}}
\subfigure[]{
\psfrag{a}[c][c][0.8]{\shortstack{source\\ plane}}
\psfrag{b}[c][c][0.8]{\shortstack{Metasurface\\ grating}}
\psfrag{e}[c][c][0.8]{$0^\text{th}$}
\psfrag{f}[c][c][0.8]{$\ell_0$}
\psfrag{x}[c][c][0.8]{$x$}
\psfrag{k}[c][c][0.8]{$\lambda_{1}-\lambda_{2}$}
\psfrag{m}[c][c][0.8]{$\lambda_{1}$}
\psfrag{n}[c][c][0.8]{$\lambda_{2}$}
\includegraphics[width=0.8\columnwidth]{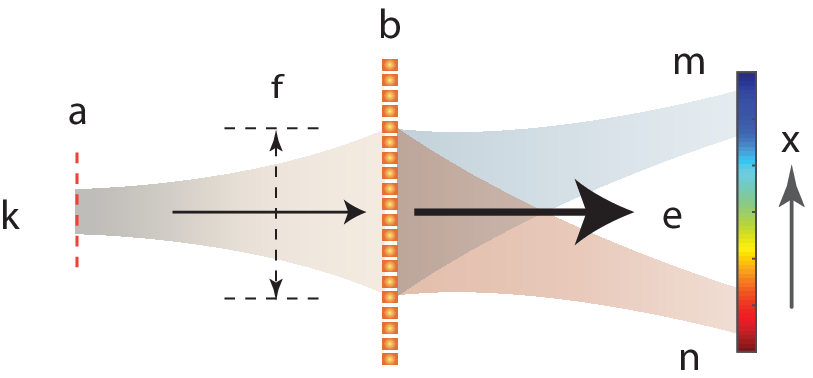}}
\caption{Fundamental wavelength discriminating functionality of a) conventional thin sinusoidal diffraction grating, and b) proposed metasurface grating.}\label{Fig:Gratings}
\end{center}
\end{figure}

\noindent The first term corresponds to the zeroth order of the grating and is independent of the wavelength. The last two terms are the $+1^\text{st}$ and $-1^\text{st}$ diffraction orders, in which light of different wavelengths propagate in different directions. The corresponding wavelength-to-space mapping is given by $x = \pm (\lambda d)/\Lambda$, and this mapping is conventionally used for wavelength separation purposes. These orders are illustrated in Fig.~\ref{Fig:Gratings}(a). Since, only one of the diffraction orders is typically used, the overall diffraction efficiency of the usable order is $(m/4)^2$ resulting in a grating efficiency always less than unity due to other spurious orders.

\begin{figure}[htbp]
\begin{center}
\psfrag{a}[c][c][0.9]{\shortstack{source\\ plane}}
\psfrag{b}[c][c][0.9]{Metasurface}
\psfrag{c}[c][c][0.9]{$\lambda/2$}
\psfrag{d}[r][c][0.7]{$\phi(\lambda_2)=-\phi_0\ell_0/2$}
\psfrag{e}[r][c][0.7]{$\phi(\lambda_2)=+\phi_0\ell_0/2$}
\psfrag{f}[c][c][0.9]{$\theta$}
\psfrag{g}[c][c][0.9]{$d$}
\psfrag{h}[c][c][0.9]{$x_0(\lambda_2)$}
\psfrag{j}[c][c][0.9]{$\ell_0$}
\psfrag{x}[c][c][0.9]{$x$}
\psfrag{k}[r][c][0.8]{$\phi(\lambda_1)=0$}
\psfrag{m}[r][c][0.8]{$\phi(\lambda_1)=0$}
\psfrag{n}[l][c][0.8]{$x_{0}(\lambda_1)=0$}
\psfrag{q}[l][c][0.8]{$\lambda_2$}
\psfrag{r}[l][c][0.8]{$\lambda_1$}
\psfrag{s}[c][c][0.8]{$\phi(x)$}
\psfrag{t}[c][c][0.8]{$\Delta\phi$}
\psfrag{z}[l][c][0.8]{$\theta(\lambda_2)$}
\includegraphics[width=\columnwidth]{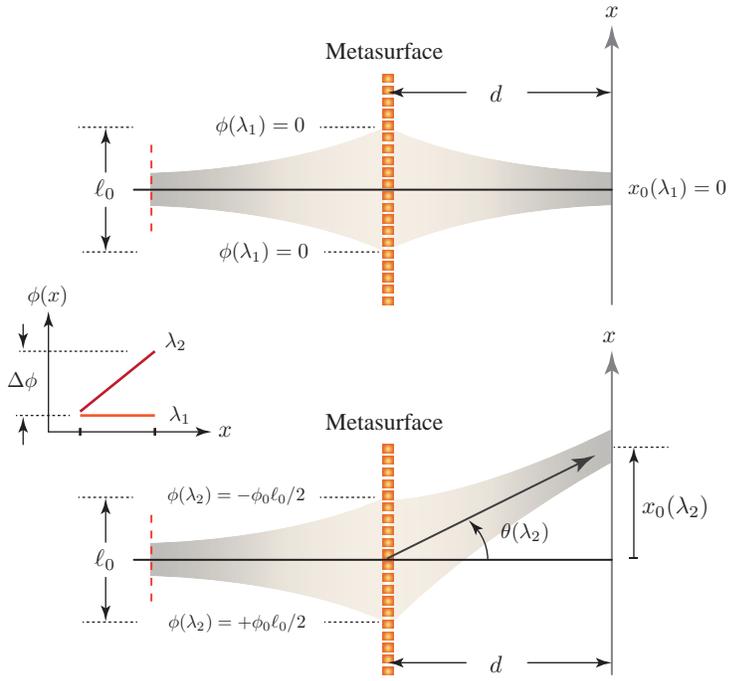}
\caption{Principle of the proposed metasurface grating where the metasurface is dispersion engineered to refract different wavelengths towards different angles. The metasurface grating is assumed to be followed by a focal lens which is not shown here for simplicity.}\label{Fig:Principle}
\end{center}
\end{figure}

\subsection{Metasurface Diffraction Grating}

Let us consider again a diffraction grating with a spatial period of $\Lambda$ in the transverse direction along $x$. The spatial frequencies associated with this spatial variation are given by $k_x = q(2\pi/\Lambda)$, where $q \in \mathcal{I}$. The wavenumber of the transmitted wave follows the free-space dispersion, $k_0^2 = k_x^2 + k_z^2$, and is thus given by

\begin{equation}
k_z = 2\pi\sqrt{\left(\frac{1}{\lambda}\right)^2 - \left(\frac{q}{\Lambda}\right)^2}.
\end{equation}

\noindent  All possible integer values of $q< \Lambda/\lambda$ results in real $k_z$ and represent the propagating diffraction orders of the grating. When $\Lambda > \lambda$, there are discrete permissible values of $q$ leading to several diffraction orders. However, when $\Lambda < \lambda$, the only diffraction order allowed by the grating corresponds to $q=0$, i.e. the zeroth order. Therefore, in order to eliminate the higher-order diffraction beams, the grating must have a sub-wavelength period structure. Metasurfaces are such structures composed of 2D periodic arrays of sub-wavelength particles and dominantly operate in the zeroth order.

Next consider a \emph{non-dispersive} metasurface\footnote{A non-dispersive device is one whose transmission phase is a linear function of frequency $\nu$ or $1/\lambda$.}, whose transmission phase is linearly tapered along $x-$axis. The transmittance function of such a metasurface is given by

\begin{equation}
T(x,y; \lambda) =  \exp\{j\phi(\lambda)x\} =  \exp\left\{j\frac{2\pi\Phi_0}{\lambda}x\right\}
\end{equation}

\noindent where $\Phi_0$ is a constant. If the metasurface is followed by a focussing lens of focal length $d$, the intensity distribution at the focal plane of the lens is then given by

\begin{equation}
|\psi(x,y,d, \lambda)| = |\psi(x,y,0; \lambda)T(x,y)h_\text{lens}(x,y)\ast h_\text{air}(x,y)|.,
\end{equation}

\noindent  where $h_\text{lens}(x,y)$ and $h_\text{air}(x,y)$ are the impulse responses of the lens and the free-space. Assume for simplicity a plane wave excitation so that $\psi(x,y,0; \lambda) = 1$. Under paraxial conditions, the output intensity can be shown to be

\begin{align}
|\psi(x,y,d; \lambda)| \propto \delta\left(x - \Phi_0d\right). \label{Eq:tiltIdeal}
\end{align}

\noindent Therefore a non-dispersive metasurface refracts the plane wave and focusses all the wavelengths to $x = \Phi_0d$. A non-dispersive metasurface, is thus incapable of wavelength discrimination.

Next, consider a \emph{dispersive} metasurface with the following property:

\begin{align}
\phi(\lambda_1, x) = 0 &\Rightarrow T(x,y; \lambda_1) = 1\\
\phi(\lambda_2, x) = \phi_0 x&\Rightarrow T(x,y; \lambda_2) =  \exp\left\{j\phi_0x\right\}
\end{align}

\noindent where $\phi_0$ is the phase gradient along the $x-$axis, defined as

\begin{equation}
\phi_0(\lambda_2) = \frac{\angle T(x,y; \lambda_2)}{dx}.
\end{equation}

\noindent The wave at $\lambda_1$ experiences an $x-$independent transmission phase, and thus following \eqref{Eq:tiltIdeal}, focusses at $x=0$, as shown in the upper half of Fig.~\ref{Fig:Principle}. On the other hand, the wave at $\lambda_2$ can be shown to be focussed at

\begin{align}
x = \left(\frac{\phi_0\lambda_2 d}{2\pi}\right), \label{Eq:tilt}
\end{align}

\noindent as shown in the lower half of Fig.~\ref{Fig:Principle}, and thus can be spatially discriminated from $\lambda_1$. Following this example, if the metasurface can be engineered to exhibit a non-zero $\phi_0(\lambda)$, different wavelengths will focus onto different $x-$co-ordinates and will form a prism-type rainbow effect on the image plane. The outgoing angle of the wave after the metasurface can be calculated using \eqref{Eq:tilt} to obtain the beam scanning law which is given by

\begin{equation}
\theta(\lambda) = \tan^{-1}\left(\frac{\phi_0\lambda}{2\pi}\right).\label{Eq:ScanLaw}
\end{equation}

Dispersion in metasurfaces is thus a necessary condition to achieve wavelength discrimination capability. 

\begin{figure}[htbp]
\begin{center}
\subfigure[]{
\psfrag{b}[c][c][0.8]{Metasurface Grating}
\psfrag{c}[c][c][0.8]{$\ell_0$}
\psfrag{e}[c][c][0.8]{$\psi(x,y, z=d)$}
\psfrag{g}[c][c][0.8]{wavelength, $\lambda$ (nm)}
\psfrag{h}[r][c][0.8]{$\lambda_r(x)$}
\psfrag{i}[c][c][0.8]{$z=0$}
\psfrag{j}[c][c][0.8]{$n$}
\psfrag{q}[c][c][0.8]{$r_1$}
\psfrag{m}[c][c][0.8]{$\psi(x,y, z=0)$}
\psfrag{p}[l][c][0.8]{$x = x_0(\lambda)$}
\psfrag{g}[c][c][0.8]{$ \Lambda \ll \lambda$}
\psfrag{x}[c][c][0.8]{$x$}
\psfrag{y}[c][c][0.8]{$y$}
\psfrag{z}[c][c][0.8]{$z$}
\psfrag{d}[c][c][0.8]{$\mathbf{m}$}
\psfrag{f}[c][c][0.8]{$r_2$}
\psfrag{k}[c][c][0.8]{$\mathbf{p}$}
\includegraphics[width=0.65\columnwidth]{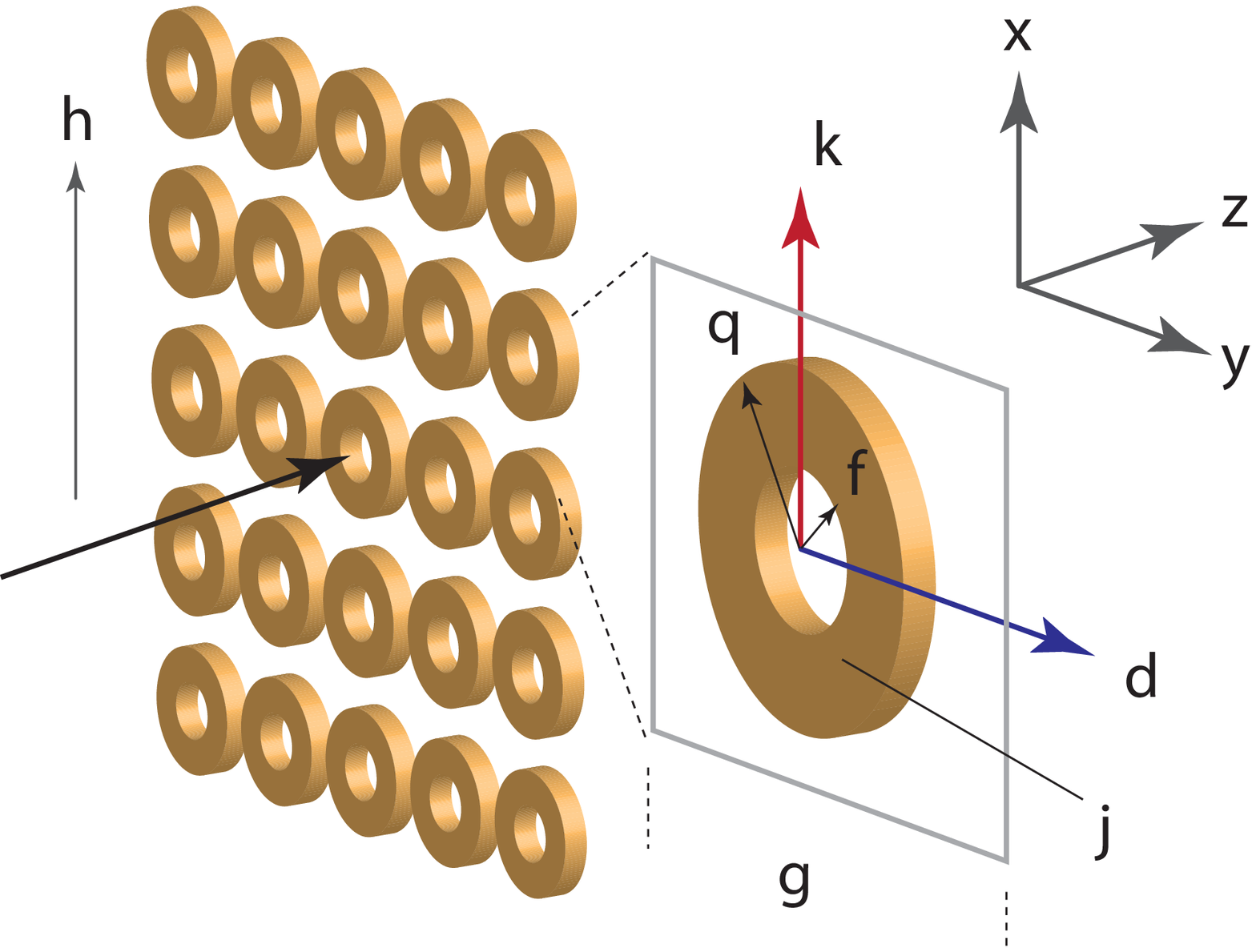}}
\subfigure[]{
\psfrag{a}[c][c][0.8]{wavelength, $\lambda$ (nm)}
\psfrag{c}[c][c][0.8]{delay, $\tau(x,y;\lambda)$ (ps)}
\psfrag{b}[c][c][0.8]{$\omega_p/(2\pi) \times 10^{-12}$ (THz)}
\psfrag{d}[l][c][0.8]{polynomial fit}
\psfrag{m}[l][c][0.8]{analytical fit}
\psfrag{e}[c][c][0.8]{resonant wavelength $\lambda_r$ (nm)}
\psfrag{n}[l][c][0.6]{$\begin{tabular}{ ccc } 
  $r_2$ & $r_1$ & $\Lambda$ \\\hline
   100 & 325 & 700 \\ 
  77 & 300 & 665 \\
  40 & 265 & 630 \\
  30 & 250 & 600
\end{tabular}$}
\includegraphics[width=0.9\columnwidth]{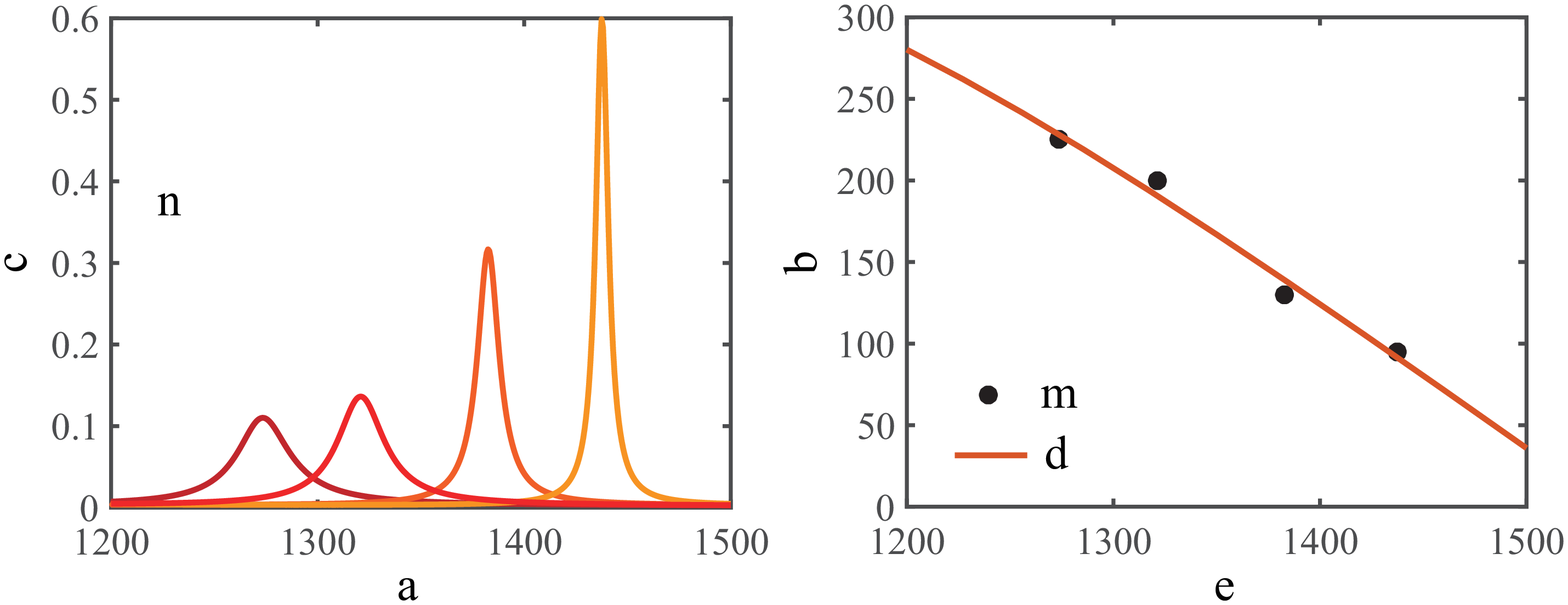}}
\caption{Dispersion engineered all-dielectric metasurfaces. a) Metasurface unit cell. b) FEM-HFSS simulated dispersion response of 4 different unit cells using periodic boundary conditions, and the extracted $\lambda_p$ vs $\lambda_r$ relation using analytical transfer function of \eqref{Eq:LorentzTF}. Here $n=3.5$, $n_0 = 1.66$ and $t=220$~nm.} \label{Fig:Setup}
\end{center}
\end{figure}

\subsection{Dispersion Engineered Grating based on Lorentz Huygen's Sources}

A metasurface is a 2D array of sub-wavelength particles. It may be modelled by surface polarizabilities, that relate the average electromagnetic fields to the induced dipole moments, i.e. $\mathbf{p} = \hat{\alpha}_eE_\text{av}$ and $\mathbf{m} = \hat{\alpha}_mH_\text{av}$, where $E_\text{av}$ and $H_\text{av}$ are the average electric and magnetic fields tangential to the metasurface, respectively, and $\hat{\alpha}_e$ and $\hat{\alpha}_m$ are the effective electric and magnetic polarizability densities per unit area \cite{Metasurface_Synthesis_Caloz}\cite{Grbic_Metasurfaces}. It is assumed that $\mathbf{p}$ and $\mathbf{m}$ are oriented perpendicular to each other and in the plane of the metasurface, a configuration that is commonly referred to as the Huygens' source. If $\hat{\alpha}_m = \eta_0^2\hat{\alpha}_e \forall\lambda$, where $\eta_0$ is the free-space impedance, it can shown that the metasurface is perfectly matched to free space and its transfer function is given by \cite{Metasurface_Synthesis_Caloz}\cite{Grbic_Metasurfaces}

\begin{equation}
T_\text{m}(\lambda) = \left[\frac{\lambda + i2\pi c\eta_0\hat{\alpha}_\text{e}(\lambda)/2}{\lambda - i2\pi c\eta_0\hat{\alpha}_\text{e}(\lambda)/2}\right].\label{Eq:TF}
\end{equation}

\noindent If Lorentz dispersion is further assumed for $\hat{\alpha}_e$, given by

\begin{equation}
\hat{\alpha}(\lambda) = \frac{\lambda_p^2}{\lambda_r}\left[\frac{\lambda_r}{\lambda_r^2 - \lambda^2} - \frac{j\pi}{2}\{\delta(\lambda + \lambda_r) + \delta(\lambda - \lambda_r)\}\right],
\end{equation}

\noindent where $\lambda_r$ is the resonant wavelength of the particles and $\lambda_p$ is the plasma wavelength, the general transfer function of \eqref{Eq:TF} becomes 

\begin{align}
T(\lambda) = \left[\frac{\lambda(\lambda_r^2 - \lambda^2) + j\pi \eta_0 c\lambda_p^2}{\lambda(\lambda_r^2 - \lambda^2) - j\pi \eta_0 c\lambda_p^2}\right] = \exp\{j\phi(\lambda)\} \label{Eq:LorentzTF}
\end{align}

\noindent noting that under these assumptions, the metasurface acts as an all-pass surface, with $|T(\lambda)| = 1\forall \lambda$. Its corresponding wavelength dependent transmission phase is thus given by

\begin{equation}
\phi(\lambda) = 2\tan^{-1}\left[\frac{\pi \eta_0 c\lambda_p^2}{\lambda(\lambda_r^2 - \lambda^2)}\right].\label{Eq:Phase}
\end{equation}

\noindent The dispersion of the metasurface is therefore controllable through the resonant wavelength parameter, $\lambda_r$. It should be noted that in a physical resonator, the resonant wavelength $\lambda_r$ and the plasma wavelength $\lambda_p$ are typically related, and $\lambda_r$ is the only free design parameter. A metasurface constructed using Lorentz-type Huygens' sources, thus provide an ideal framework to engineer the dispersion of the metasurface, while ensuring a perfect match in a broad bandwidth.

\section{Metasurface Grating Demonstration}

A Lorentz metasurface is typically implemented using a 2D array of all-dielectric resonators in the form of cylindrical nanodisks, which naturally provide the Huygens' source configuration of orthogonal $\mathbf{p}$ and $\mathbf{m}$, considered herein, providing the required transfer function of \eqref{Eq:LorentzTF} \cite{AllDieelctricMTMS}\cite{AllDielectricKivshar}. Fig.~\ref{Fig:Setup}(a) shows an example of an all-dielectric metasurface, which is particularly suitable for engineering the metasurface dispersion. The unit cell of size $\Lambda< \lambda$, is composed of a nanodisk of radius $r_1$ and $r_2$ made of a material with a refractive index $n$ which could be buried in a host dielectric of refractive index $n_0$. Radii $r_1$, $r_2$ and the unit cell size $\Lambda$, control the resonant wavelength $\lambda_r$ of the metasurface, for a given index $n$ and disk thickness $t$, and thus can be designed to engineer the metasurface dispersion.

Fig.~\ref{Fig:Setup}(b) shows the typical dispersion (delay) response, obtained using periodic boundary conditions in FEM-HFSS, for varying physical parameters of the metasurface unit cell formed using silicon nanodisks with $n=3.5$ buried in a host dielectric of refractive index $n_0 = 1.66$ of thickness $t=220$~nm. The bandwidth chosen here for demonstration is between 1200-1500~nm. As can be seen, the resonant wavelength, where the delay is maximum, can be conveniently varied over a large bandwidth, ensuring a near perfect match in the entire bandwidth ($20\log|T(\lambda)| < -10$~dB, not shown here). It is also noticed that as $\lambda_r$ increases, the delay swing increases resulting in a stronger dispersion. 

The transmission response of each nano-disk unit cell is then analytically fitted using the Lorentz transfer function of \eqref{Eq:LorentzTF} to extract the corresponding plasma wavelength $\lambda_p$. The resulting plot of $\lambda_p$ vs $\lambda_r$ is also shown in Fig.~\ref{Fig:Setup}(b) which is then used to obtain a continuous distribution using polynomial fitting. This polynomial function thus acts as a look-up function to realize a continuous sweep of $\lambda_r$ across the entire bandwidth.

Next, a metasurface is dispersion engineered by sweeping $\lambda_r$ linearly along the $x-$axis, to create a non-zero phase gradient $\phi_0$, following the relationship:

\begin{equation}
\lambda_r(x, y) = \lambda_\text{min} + (x + \ell_0/2)\frac{\lambda_\text{max} - \lambda_\text{min}}{\ell_0}.
\end{equation}

\begin{figure}[htbp]
\begin{center}
\subfigure[]{
\psfrag{a}[c][c][0.8]{$x$ (mm)}
\psfrag{b}[c][c][0.8]{$x~(\mu\text{m})$}
\psfrag{c}[c][c][0.8]{phase, $\angle T(x,y; \lambda)/\pi$ (rad)}
\psfrag{d}[c][c][0.8]{$\lambda$}
\psfrag{e}[c][c][0.8]{$2\lambda_\text{BW}$}
\psfrag{f}[c][c][0.8]{wavelength, $\lambda$ (nm)}
\includegraphics[width=\columnwidth]{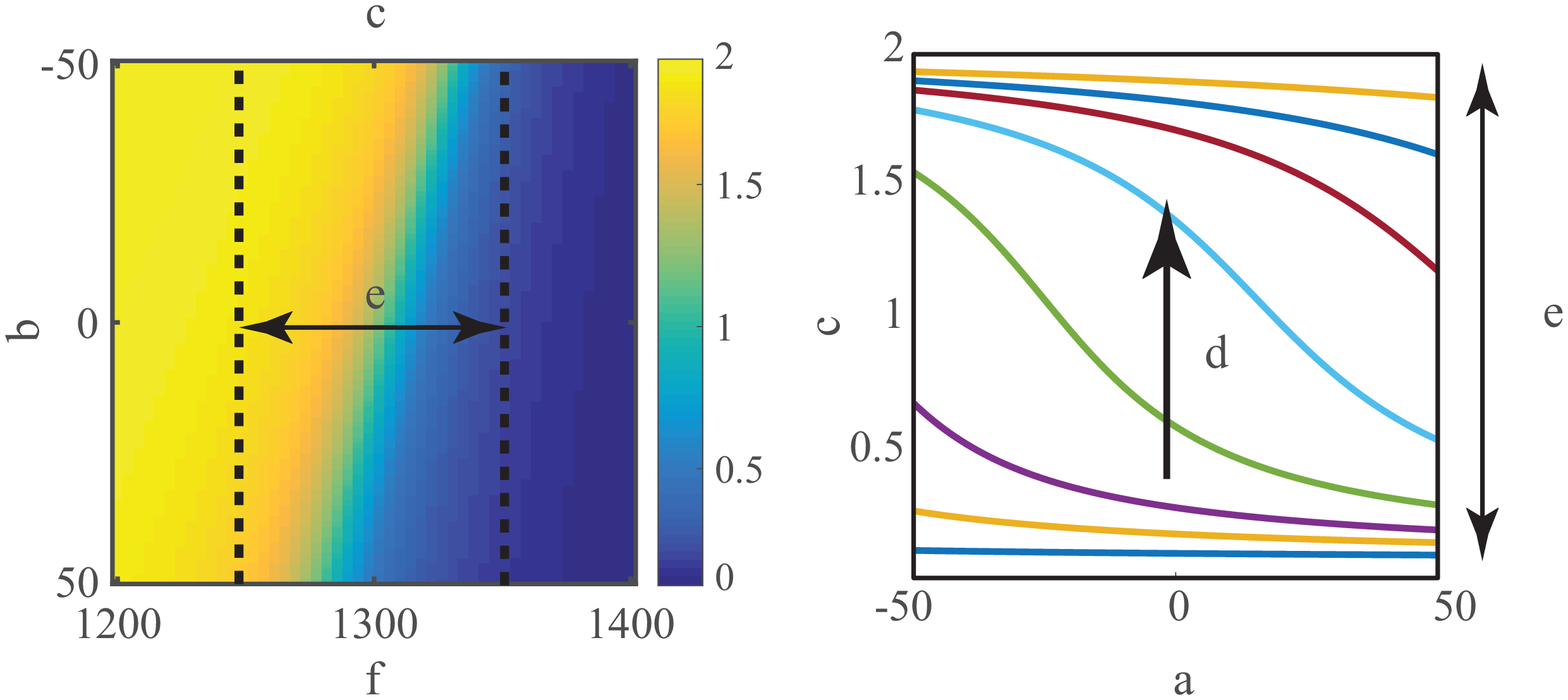}}
\subfigure[]{
\psfrag{a}[c][c][0.8]{wavelength, $\lambda$ (nm)}
\psfrag{b}[c][c][0.8]{Scan angle $\theta$ (deg)}
\psfrag{x}[c][c][0.8]{$x$}
\psfrag{d}[c][c][0.8]{$\theta(\lambda)$}
\psfrag{c}[l][c][0.8]{$\displaystyle{\phi_0(\lambda) = \frac{d\angle T}{dx}}$}
\psfrag{e}[c][c][0.8]{Band~\#1}
\psfrag{f}[c][c][0.8]{Band~\#2}
\psfrag{g}[c][c][0.8]{$\lambda_\text{BW}$}
\includegraphics[width=0.8\columnwidth]{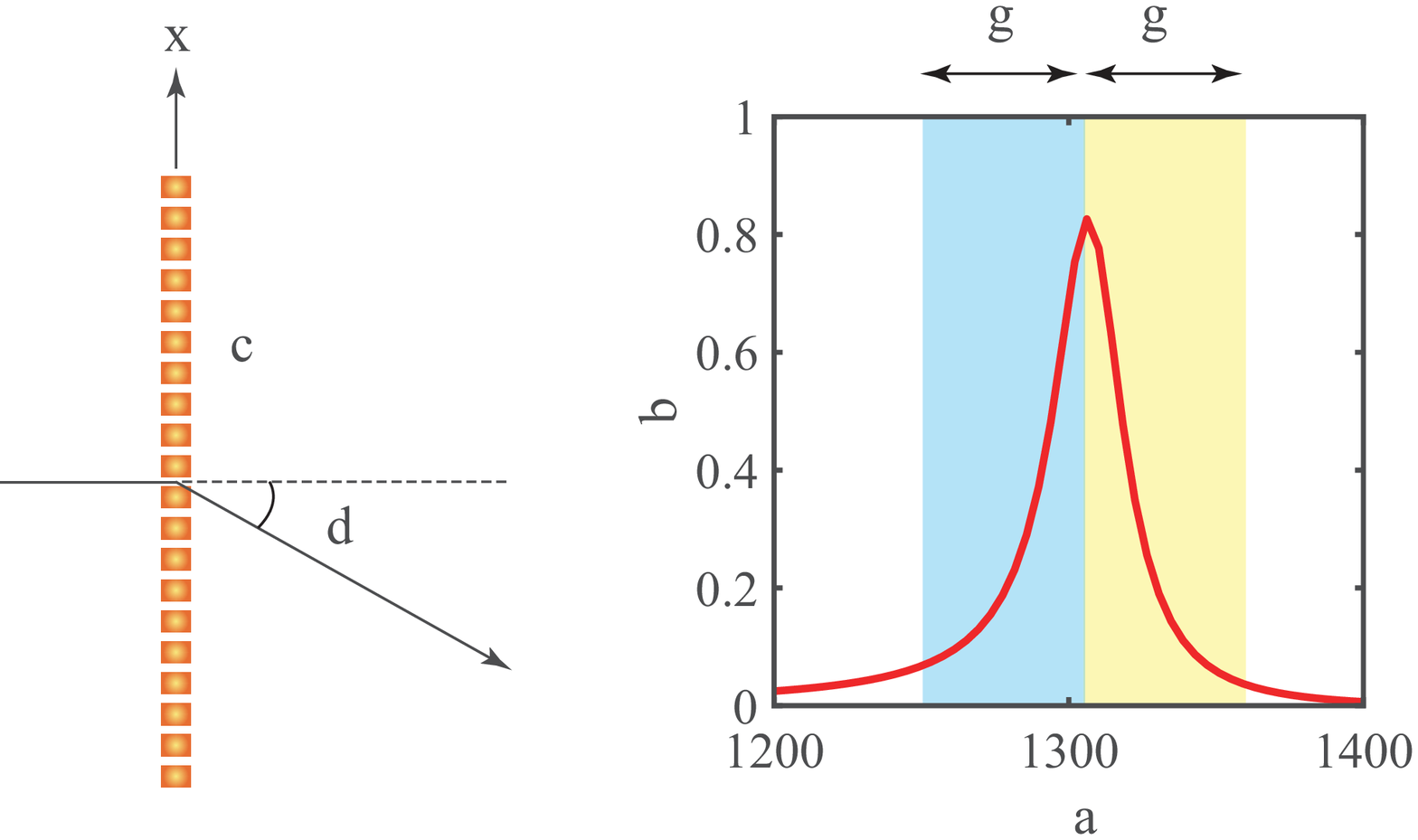}}
\caption{Metasurface grating characteristics. a) Metasurface transmission phase as a function of wavelength along the $x-$axis of the metasurface. b) Wavelength scanning law. Here $\lambda_\text{max} = 1325$~nm and $\lambda_\text{min} = 1275$~nm and $\ell_0 = 100~\mu$m.}\label{Fig:PhaseMap}
\end{center}
\end{figure}

\noindent The typical transmission phase distribution of the metasurface is shown in Fig.~\ref{Fig:PhaseMap}(a) as function of the wavelength and the spatial co-ordinate $x$. The metasurface is $\ell_0^2 = 100~\mu\text{m}\times 100~\mu\text{m}$ in size. As clearly seen in the figure, the phase gradient $\phi_0$ is a function of wavelength. Since $\phi_0$ is also a function of $x$, to obtain the scan law $\theta(\lambda)$ of the metasurface, $\phi_0$ is evaluated at the centre of the metasurface at $x=0$. Fig.~\ref{Fig:PhaseMap}(b) shows the obtained scanning law of this metasurface. The scanning law is strongly nonlinear and thus has varying wavelength resolution depending on the wavelength, which becomes finer near ~1300~nm, i.e. large change in $\theta$ for a small change in $\lambda$. This is due to the fact that the phase gradient parameter depends on wavelength as opposed to a desired constant value. Further, the scan possesses a symmetry around 1300~nm enabling two separate bands of operation with identical wavelength resolutions. The metasurface can thus be either operated in 1200-1300~nm band or 1300-1400~nm band. In each band, the angular dispersion of the grating $\approx 0.8^\circ/50$~nm.

\begin{figure*}[htbp]
\begin{center}
\psfrag{a}[c][c][0.8]{$z$ (cm)}
\psfrag{b}[c][c][0.8]{$x(\mu\text{m}$)}
\psfrag{c}[c][c][0.8]{$\lambda = 1260$~nm}
\psfrag{d}[c][c][0.8]{$\lambda = 1270$~nm}
\psfrag{e}[c][c][0.8]{$\lambda = 1280$~nm}
\psfrag{f}[c][c][0.8]{$\lambda = 1290$~nm}
\psfrag{g}[c][c][0.8]{$\lambda= 1300$~nm}
\includegraphics[width=1.9\columnwidth]{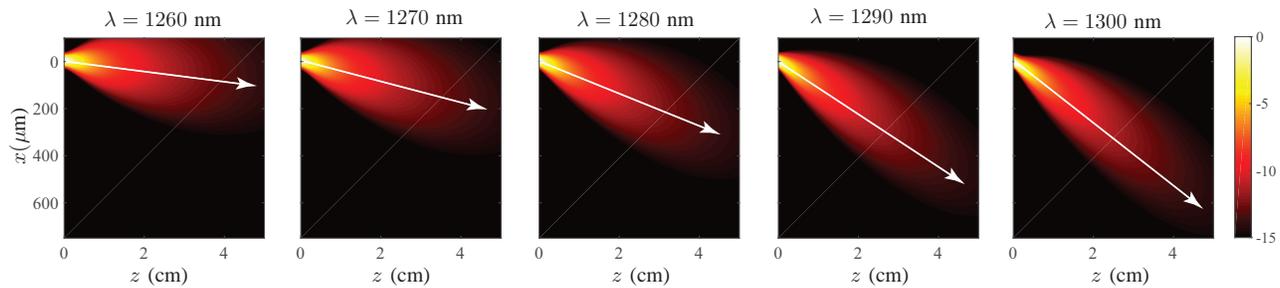}
\caption{The analytical field distribution in the $x-z$ plane when a gaussian beam of width $w_0$ is incident on the dispersion engineered metasurface of Fig.~\ref{Fig:PhaseMap}, for various different wavelengths. The Gaussian beam assumed is $\psi(x,y,0) = \exp\{-(x^2 + y^2)/2w_0^2\}$.}\label{Fig:Results}
\end{center}
\end{figure*}

Figure.~\ref{Fig:Results} shows the computed field distribution at different wavelengths when the metasurface of Fig.~\ref{Fig:PhaseMap} is incident with a Gaussian beam. The fields are computed using analytical Fourier propagation such that $\psi(x,y,z; \lambda) = \psi(x,y,z; \lambda)T(x,y,z; \lambda)\ast h_\text{air}(x,y,z; \lambda)$. As expected, different wavelengths propagate in different direction after hitting the metasurface following the scanning law, and thus form a spectral rainbow at the output image planes. This demonstrates the wavelength separation capability of the designed all-dielectric metasurface grating. 

\section{Features and Benefits}

\begin{enumerate}
\item The metasurface grating can be seen as a flat prism, which is engineered on an "atomic" scale to achieve wavelength separation with 100\% transmission. Dispersion engineering at the unit scale dimensions, provides a highly flexible grating design, adaptable for specified bandwidths and wavelength resolution requirements. 
\item Similar to optical prisms, the proposed metasurface grating operates in the fundamental zeroth order diffraction mode. Compared to conventional diffraction gratings, no energy is wasted in the higher orders in the met surface gratings, thereby making it possible to theoretically achieve a maximum efficiency of 100\%. 
\item The all-dielectric metasurface implementation based on Huygen's sources ensures a perfect match to free-space which is a desirable feature for such a transmission-mode device. 
\item Compared to the typically used plasmonic unit cells, the dielectric unit cells employed in the metasurface design herein, is ideal for optical frequencies offering low dissipation losses, to help maximize the grating efficiency. 
\end{enumerate}

\section{Conclusions}

A single-order transmission diffraction grating based on dispersion engineered all-dielectric metasurfaces has been proposed. It's wavelength discriminating properties have been theoretically described and its operation has been confirmed using numerical simulations. The metasurface has been designed using all-dielectric unit cell, which emulates a Huygens' source configuration to achieve a perfect match to free-space in broad bandwidth. Using a holey dielectric nanodisk structure as the unit cells, the resonant wavelength is varied across the metasurface to engineer the wavelength dependent spatial phase gradient, required for achieving the wavelength discrimination. Finally, a numerical demonstration has been provided to achieve an angular dispersion of about $0.8^\circ/50$~nm within 1250-1300~nm bandwidth. Its experimental validation is currently under progress and will be reported elsewhere.
 
\section*{Acknowledgement}

The author would like to thank Dr. Fangxin Li, from Synaptive Medicals, Toronto, for fruitful discussions related to diffraction gratings and state-of-the art spectrum analyzers.

\bibliographystyle{IEEEtran}
\bibliography{2016_Gupta_SingleOrderGrating_JOSA-A}

%

\end{document}